\documentclass[12pt,preprint]{aastex}
\usepackage{lscape}
\usepackage{natbib}
\usepackage{multicol}

\newcommand{\lsim}{\raisebox{-0.3ex}{\mbox{$\stackrel{<}{_\sim} \,$}}}
\slugcomment{Version from \today}

\begin{document}

\title{Direct Detection of a (Proto)Binary-Disk System in IRAS 20126+4104}
\author{T.K. Sridharan\altaffilmark{1}}
\affil{Harvard-Smithsonian Center for Astrophysics, 60 Garden Street, MS 78, Cambridge, MA 02138, USA.}
\author{S.J. Williams\altaffilmark{1}, G.A. Fuller\altaffilmark{1}}
\affil{Physics Department, UMIST, P.O. Box 88, Manchester M60 1QD, UK}
\altaffiltext{1}{emails: tksridha@cfa.harvard.edu, Stewart.Williams@manchester.ac.uk, Gary.Fuller@manchester.ac.uk}

\begin{abstract}

We report the direct detection of a binary/disk system towards the
high-mass (proto)stellar object IRAS20126+4104 at infrared wavengths. 
The presence of a 
multiple system had been indicated by the precession of the outflow and 
the double jet system detected earlier at cm-wavelengths. Our new K, L$^{\prime}$ \& 
M$^{\prime}$ band infrared images obtained with the UKIRT under exceptional seeing 
conditions on Mauna Kea are able to resolve the central source for the first time, 
and we identify two objects separated by $\sim$ 0.5$^{\prime\prime}$ (850 AU).  The K and L$^{\prime}$ 
images also uncover features characteristic of a nearly edge-on disk, 
similar to many low mass protostars with disks: two emission regions oriented 
along an outflow axis and separated by a dark lane. The peaks of the 
L$^{\prime}$ \& M$^{\prime}$ band and mm-wavelength emission are on the dark lane, 
presumably locating the primary young star. The thickness of the disk is measured to be
$\sim$ 850 AU for radii $\lsim$ 1000 AU. Approximate limits on 
the NIR magnitudes of the two young stars indicate a high-mass system, although with 
much uncertainty.  
These results are a demonstration of the high-mass nature of the system,
and the similarities of the star-formation process in the low-mass and 
high-mass regimes viz. the presence of a disk-accretion stage. The companion is 
located along the dark lane, consistent with it being in the equatorial/disk plane, 
indicating a disk-accretion setting for massive, multiple, star-formation.

\end{abstract}


\section {Introduction}

The study of high-mass star-formation has attracted much recent attention.
Molinari et al (1996, 2002) and Sridharan et al (2002)  have reported 
systematic studies of candidate high-mass proto-stellar objects (HMPOs),
leading to a number of follow-ups.  A key unresolved question is: how similar are 
the high- and low-mass star-formation processes? Do high-mass stars go through a 
disk-accretion phase, which is a central idea in the current, widely accepted
scenario for low-mass star-formation? Sufficient observational data is beginning
to be available for the HMPOs to answer these questions. The outflow phase has 
been shown to be as prevalent towards the HMPOs as towards lower-mass objects, and 
by implication an accretion disk can be inferred (Zhang et al 2001, Beuther et 
al 2002b). While there is good evidence for rotating disks of radius $\sim$ several 1000 AU 
(Cesaroni et al 1997, 2005, Zhang, Hunter \& Sridharan, 1998, Beltran et al 2005), 
outflows presumably require a compact disk and direct evidence for their presence 
is rare (Patel et al., 2005). Compact disks have been proposed previously, 
although not directly imaged (Cesaroni et al 1997 \& 1999; Shepherd \& Kurtz 1999, 
Shepherd, Claussen \& Kurtz, 2001). 

Among HMPO candidates, IRAS20126+4104 is particularly well studied 
(Cesaroni et al 1997, 1999, 2005, Zhang et al 1999, Yao et al, 2000, Shepherd et al 2000). It has a luminosity of 
$\sim$ 10$^4$ L$_{\odot}$ at a distance of 1.7 kpc. Recent observations
suggest a mass of $\sim$ 7 M$_{\odot}$ (Cesaroni et al 2005), significantly lower than 
previous estimates of 20-24 M$_{\odot}$ (Cesaroni et al 1999, Zhang, Hunter
\& Sridharan, 1998).
It harbours water, hydroxyl and 
methanol masers (Tofani et al 1995, Minier et al 2001, Edris et al 2005) and
an outflow/jet system traced in CO, SiO, H$_2$ and cm-wavelength continuum 
emission. It has a rotating disk of radius $\sim$ 5000 AU, 
traced in ammonia and CS (Zhang, Hunter \& Sridharan 1998; Cesaroni et al 2005) and a possible 
smaller $\sim$ 1000 AU radius disk inferred from interferometric 
CH$_3$CN measurements (Cesaroni et al 1999). 
Sensitive cm-wavelength 
continuum images have uncovered emission south of the central 
object suggesting the presence of a companion (Hofner et al 1999),
also indicated by the precession of the H$_2$
jet from this source (Shepherd et al 2000, Cesaroni et al 2005). We note that 
IRAS20126+4104 is a member of a 
large sample of HMPO candidates we have been studying systematically (Sridharan et al 2002, 
Beuther et al 2002a, b, Williams, Fuller \& Sridharan 2004, 2005).


Here, we report the infrared detection of a compact disk/binary system
in IRAS20126+4104, consisting of a small scale circumstellar disk and
a secondary source.

\section {Observations}

The observations were carried out during 15-18 Aug, 2000. We imaged 
IRAS20126+4104 using the near-infrared cameras UFTI (1-2.5 $\mu$m) and TUFTI/IRCAM 
(1-5 $\mu$m) on the United Kingdom Infrared Telescope
(UKIRT).\footnote{The UKIRT is operated by
   the Joint Astronomy Centre on behalf of the U.K. Particle Physics and Astronomy Research
Council}
TUFTI is a 256$\times$256 
InSb Array with a field of view 
of 20.7$^{\prime\prime}$ and 0.08$^{\prime\prime}$ pixels. UFTI has a 
1024$\times$1024 HgCdTe Array with 
0.09$^{\prime\prime}$ pixels and a field of view of 93$^{\prime\prime}$.
The seeing during these measurements was excellent, at  $\sim$ 0.3$^{\prime\prime}$
(K band). Observations were
made with broadband K, L$^{\prime}$, and M$^{\prime}$ filters (central wavelengths of 2.2, 3.8 \& 4.7$\mu$m).
The K band
observations employed a 11$^{\prime\prime}$ jitter and the L$^{\prime}$ and M$^{\prime}$ 
observations
used 8.5$^{\prime\prime}$, 7.5$^{\prime\prime}$ and 
6.5$^{\prime\prime}$ jitters and a 2$^{\prime\prime}$
jitter with a 30$^{\prime\prime}$ chop. Standard data reduction steps were used to obtain
the final images. Several calibration stars were observed 
for reliable photometry.

We used K-band objects from the 2MASS survey to fit separate plate 
solutions for the K and L$^{\prime}$ images. The formal errors for the plate
solutions are 0.15$^{\prime\prime}$ rms, consistent with the 2MASS
astrometric accuracy\footnote{2MASS Explanatory Supplement, 
http://ipac.caltech.edu/2mass/releases/allsky/doc/expl-sup.html}.
The error on the relative registration is 0.1$^{\prime\prime}$ rms,
 measured using the 
positions of the same objects in the independently registered K and L$^{\prime}$ 
images.
Since the M$^{\prime}$ image does not include any 
reference objects, we cross-correlated it with the L$^{\prime}$ image to 
find the 
best position offset. We estimate the absolute registration uncertainty  
to be $\sim0.15''$.

\section {Results and Discussion}

As shown in Figure 1, we detect two, previously unknown, faint lobes of emission 
in the K-band, K-SE \& K-NW, near the center of the IRAS20126+4104 outflow region. Previous images
only showed the bi-polar features on larger $\sim$ 10$^{\prime\prime}$ scales 
oriented at about the same position angle as the molecular and
the cm-wavelength outflows (Cesaroni et al 1997, Ayala et al 1998, Shepherd et al, 
2000). The new emission objects are separated by 
a dark lane with the line joining them oriented NW-SE, similar to 
the larger scale bi-polar features.
Figure 2 (left panel) shows the L$^{\prime}$ 
and K emission in the 
central 2$^{\prime\prime}$, where
the center of the dark lane is seen to nearly coincide with the location 
of the mm-wavelegth emission from interferometric images.
The L$^{\prime}$ emission over 
the region lies close to the dark lane seen in the K band. The peak of
this central emission is nearly centered on the dark lane,
within registration errors. Emission is also seen towards the K-band lobes. In addition,
there is a point source like object seen $\sim$ 0.5$^{\prime\prime}$ southwest 
of the center. This second object is at one 
end of, and in line with, the dark lane. The center panel in Figure 2 shows 
the M$^{\prime}$ emission 
contours superposed on the K image. The  M$^{\prime}$ emission has 
a similar morphology as that in the L$^{\prime}$ band, but with the central object more prominent.  
The correspondence between the emission in the L$^{\prime}$ and the M$^{\prime}$
bands is shown in the right panel.  
The source detected at L$'$ to the southwest of the peak is also seen in the M$'$ image,
confirming its reality. This also validates our method of 
cross-correlating L$^{\prime}$ and M$^{\prime}$ band images for registration,
which is sensitive only to the strongest emission features in the images.

Bi-polar infrared morphologies similar to the ones found in our data are
frequently seen in near infrared observations of low mass young
stars, for example the HST images 
of T Tauri stars by Padgett et al. (1999).  Towards the low 
mass objects these morphologies have been interpreted as 
due to the presence of nearly edge-on circumstellar disks.
Therefore we suggest that our images represent the first detection of a small 
scale disk towards a young high-mass star, as a silhoutte at NIR wavelengths. 
The width of the dark lane, which corresponds
to the thickness of the disk is $\sim$ 0.5$^{\prime\prime}$ and the diameter is less certain 
at $\sim$ 1$^{\prime\prime}$. The implied linear dimensions of the disk are 
$\sim 850$ AU thickness, for radii \lsim 1000 AU.  This is comparable to 
the size of the disk 
inferred from the CH$_3$CN data of Cesaroni et al (1999). 
These sizes are significantly larger than the small-scale disks seen towards 
low-mass star-forming regions as may be expected for higher-mass proto-stars.

The presence of the second source makes this object more interesting. 
As seen in Figs 2-4, this second object is 
located on the same side of the central object as the cm-wavelength jet of 
Hofner et al (1999; shown by a cross mark to the south). Given the closeness of 
the two  L$^{\prime}$ \& M$^{\prime}$ band objects, ($\sim$ 0.5$^{\prime\prime}$ 
= 850 AU), we may be seeing a young 
high-mass binary system. We suggest that the second object to the southwest is 
the companion responsible for the precessing H$_2$ emission outflow 
(Shepherd et al 2000, Cesaroni et al 2005). It is noteworthy that the companion is in 
line with the disk. It is probably close to the equatorial plane, although due to
unresolved projection effects, this cannot be said with certainty. We point 
out that in the CH$_3$CN disk inferred by Cesaroni et al (1999), the emission 
peaks appear to cut off  beyond the location of the companion. 
The location of the companion does not match the location of the cm-jet, which 
raises the possibility of a triple system.

\subsection {Extinction Estimates}

We now make some very simple and approximate estimates of the extinction due to 
the envelope surrounding the disk-binary system and that due to the disk, separately. 
We use the observed colours of the binary and the bi-polar features along with the
interstellar reddening law from Koorneef (1983) (in particular, $E_{L-M}/A_V=0.019$ \& 
$E_{K-L}/A_V=0.045$) for this purpose.  For all stellar objects, intrinsic
K-L$^{\prime}$ and L$^{\prime}$-M$^{\prime}$ 
colours of $\sim$ 0 are assumed. Two cases, {\it scattered} and {\it direct}, are considered,
where line of sight reddening occurs with and without preceding 
Rayleigh scattering ($\lambda^{-4}$ dependence; K-L$^{\prime}=-2.4$ \&  L$^{\prime}$-M$^{\prime}=-1.0$).

The magnitudes and colours, from aperture photometry, for the whole region, the central lobes and 
the binary 
components (central and companion) are presented in Table 1. 
For the K band, using the fluxes of the H$_2$ knots (Cesaroni et al 1997) and spectra from
Ayala et al (1998), a line emission contamination of a few percent or $<$ 0.1 magnitude is 
estimated.  Contamination in the L$^{\prime}$ and M$^{\prime}$ bands are not considered. 
This is reasonable for the L$^{\prime}$ band due to its large bandwidth. 
The integrated emission over the 
whole region, excluding obvious point-like sources, is in agreement with the K band flux reported 
by Cesaroni et al (1997). The colour of this emission is consistent with light from main sequence stars, 
implying  120 \& 116 magnitudes of $A_V$ (mean 118) for the {\it scattered} case, and $\sim$ 65 
magnitude for the {\it direct}
case. Therefore, an $A_V$ of 65 - 118  is considered to be the extinction due to the envelope. 

For the central lobes, the extinctions ({\it scattered} case) implied  are 113 and 127 (mean 120) from K-L$^{\prime}$ colours,
similar to the upper limit estimate for the $\sim$ 10$^{\prime\prime}$ features. Since the emission 
from central and companion objects and the bi-polar lobes cannot be well seperated, 
the L$^{\prime}$ magnitudes are lower limits (the lobes may be fainter), and the extinctions, 
upper limits. In summary, the envelope extinction is in the range 65 - 120.

Turning to the binary itself, assuming $\sim$ 0 for the intrinsic colours of the stars, 
the limits on the observed 
K-L$^{\prime}$ colours imply 
a lower limit (not detected in the K band) to $A_V$ towards                  
the central and the companion objects of $\sim$ 96 \& 91, for the {\it direct} case
and 149 \& 144 for the {\it scattered} case. From the L$^{\prime}$$-$M$^{\prime}$ colours,
 the values are 168 \& 148 and 221 \& 200, for the two cases, 
consistent with the lower limits from K-L$^{\prime}$ colours.  The ranges 168-221 and 148-200 
will be used for the two objects for further discussion. This translates to a H$_2$
column density of $\sim$ 2$\times$10$^{23}$ cm$^{-2}$ ($N_{H_2}/A_v = 0.94 \times 10^{21}$; Frerking, Langer \& Wilson, 1982). 
In comparsion, Beuther et al (2002a) estimated an extinction of 553 magnitudes for the dust core (11$''$ beam),
or 276 magnitude up to the center. 

It is interesting that the 
extinction towards the companion is less than that towards the central object: 
we speculate that this may be due to a gap cleared by the companion outside of which the material 
may be less dense.  As noted before, the CH$_3$CN emission peaks in Cesaroni et al (1999), which 
trace high densities, are distributed within the area bounded by the companion which supports such 
a picture. 

\subsection{Magnitudes and Masses}
These measurements allow an estimate of the range of masses for the disk and NIR magnitudes 
for the embedded stellar objects.  With an envelope extinction of $\sim$ 65-120, the excess extinctions
towards the two objects are in the range $A_V \sim$ 48-166 and  $\sim$ 28-135. This extinction may be 
attributed to the disk and translates to $N_{H_2} \sim 10^{23}$ cm$^{-2}$. For a 1000 AU $\times$ 2000 AU 
disk, a mass of $\sim$ 0.1 M$_{\odot}$ is indicated.  
This is quite uncertain due to the 
many assumptions, and our inablility to reliably separate the contributions to 
the measured magnitudes from the two embedded young stars and the bi-polar lobes. 
Additional uncertainties are due to part of the L$^{\prime}$ and M$^{\prime}$ band emission coming from hot dust in the disk which 
would change the colours, and the mass of such a component is also not considered. 
Flaring of the disk and its inclination to the line of sight, which would increase the mass estimates,  are
not considered. Therefore, we beleive 
0.1 M$_{\odot}$ to be an approximate lower limit to the mass within 1000 AU from the central star. 
In comparsion, Cesaroni et al (2005) estimate the mass of the disk within 
5000 AU from the star to be 1-4 M$_{\odot}$. They estimated peak H$_2$ column
densities for the disk of 1.5-3.4$\times$10$^{22}$ cm$^{-2}$ from C$^{34}$S
measurements (using numbers listed in their Table 4)
compared with our $\sim$10$^{23}$ cm$^{-2}$. We note
that our images do not constrain the radius of the disk - it can be
larger than $\sim$1000 AU, in which case our mass estimate would go up.
Also, observations such as ours selectively probe the lowest 
extinction paths through which photons escape, leading again to lower 
estimates for extinctions and masses. We note that Fuller, Zijlstra \& Williams (2001) estimated a similar low mass for a proposed NIR
disk in the high-mass jet object IRAS18556+0136.

What are the masses implied for the two young stars? Assuming main sequence objects and total extinctions 
of $A_V \sim$ 168-221 and $\sim$ 148-200 for the two, the extinction corrected L$^{\prime}$ magnitudes are: 
3.6 - 1.3 and 4.7 - 2.4 ($A_L/A_V = 0.045$; from Koorneef, 1983).  With a distance modulus of 11.2, 
the absolute L$^{\prime}$ magnitudes implied 
are: $-$7.6 - $-$9.9 \& $-$6.5 - $-$8.8 corresponding to very early O spectral types - O4 and 
earlier (Wegner 2000; Serabyn, Shupe \& Figer, 1998). For the primary object, 
Cesaroni et al (2005) derived a mass of $\sim$ 7 M$_{\odot}$ using the 
Keplerian velocity field they observed, singnificantly changing
earlier estimates of 20-24 M$_{\odot}$ and implied ZAMS spectral types of O8-O9 (Cesaroni et al 1999, Zhang, Hunter
\& Sridharan, 1998). Again, we emphasize that our estimates are to be considered approximate - 
in addition to the previously
mentioned uncertainties, any emission in the L$^{\prime}$ and M$^{\prime}$ bands from the disk would lower the magnitudes 
estimated for the stars -  and it is almost certain that the stars are not as early types as implied.  
Also, the main sequence assumption is questionable and the absolute NIR magnitudes of the brightest 
main sequence stars remain poorly characterized. The companion is probably only $\sim$ 1 magnitude
fainter, which makes it significantly less massive, but comparable. A significant mass for the 
companion will appear to be in conflict with
the recent observations of Keplerian velocity profile (Cesaroni et al 2005, Edris et al 2005). However, it is not clear
if the Keplerian profile holds good close to and inside  the 0.5$^{\prime\prime}$ separation of the companion.
Higher resolution observations are critical to resolving the disagreements. Nevertheless, our numbers do indicate 
that we are indeed dealing with a high-mass system.

Given the marginal linear resolution of the images, we have refrained from detailed modelling 
which should wait until higher resolution images are obtained with adaptive optics. Fortunately, 
there is a bright star $\sim$ 12$^{\prime\prime}$ from the center which makes IRAS20126+4104 an 
ideal candidate for such studies.

\section {Conclusions}

We presented high resolution NIR images of IRAS20126+4104. The K-band image 
shows a dark lane separating two, new, 1$''$-scale bi-polar emission features, with 
the emission peaks in the longer wavelength L$^{\prime}$,  M$^{\prime}$ and
mm-wave bands lying close to the dark lane. This is interpreted as due to
a nearly edge-on disk of 850 AU thickness for radii \lsim 1000 AU,
imaged and measured for the first time at infrared wavelengths.  A new, point-like, second 
object, found to the southwest of the central object, 
and in line with the disk,  is interpreted as a binary companion with a 
separation of $\sim$ 1000 AU.  Estimates of the NIR magnitudes of the 
objects, although very approximate, indicate a 
high-mass system.  Through their striking resemblence to 
similar low-mass systems, the new images demonstrate that the processes involved in the 
formation of low and high mass stars may have much in common: viz., multiple star formation 
in an accretion disk setting.  Further higher resolution studies 
will be extremely rewarding in modelling the system and therefore to address a number of 
issues related to high-mass star-formation.

\acknowledgements
We thank the referee, Riccardo Cesaroni, for careful comments which significantly improved the 
paper.

\newpage
\includegraphics[angle=-90,width=15cm]{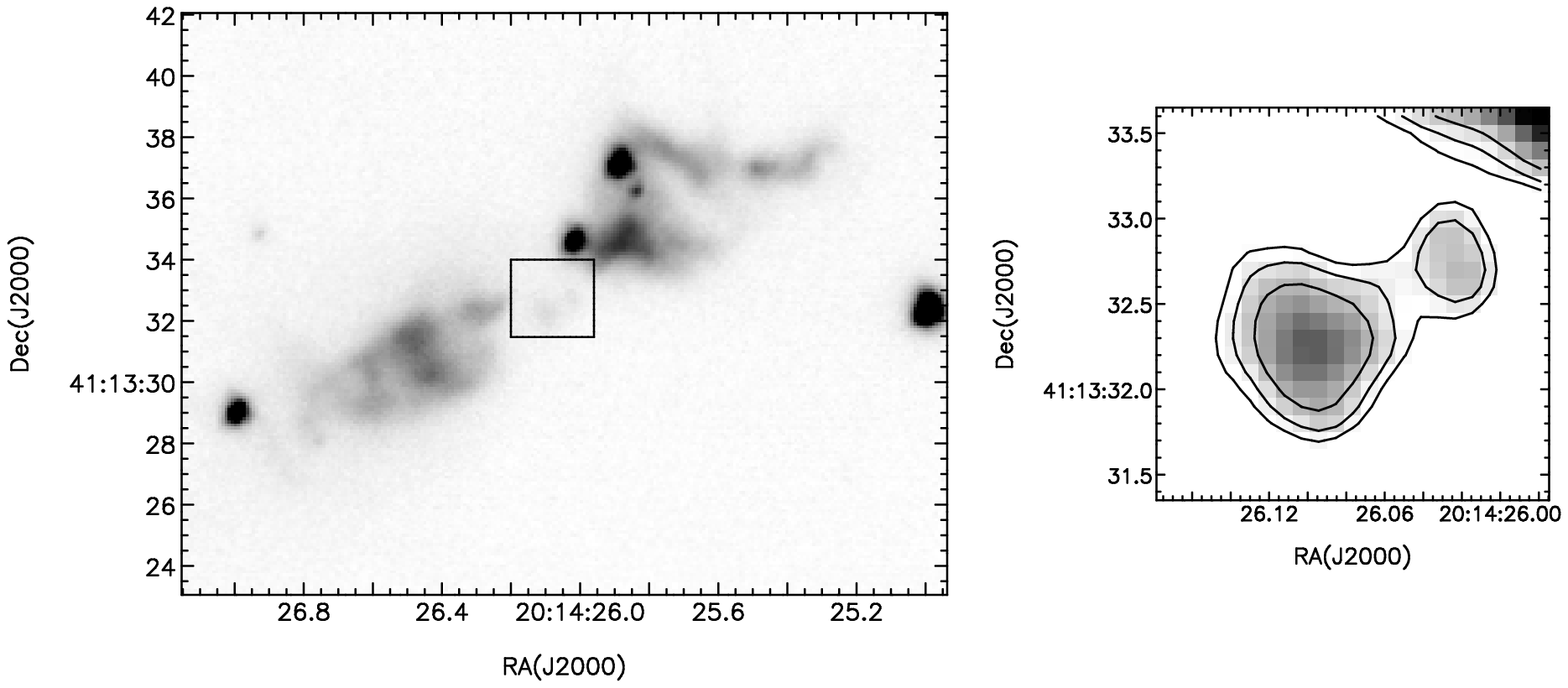}
\figcaption{\small K-band image of IRAS20126+4104 (J2000 co-ordinates). The central region, shown
as a box, is enlarged in the right panel, where two faint emission 
lobes are seen.}

\newpage
\includegraphics[angle=-90,width=15cm]{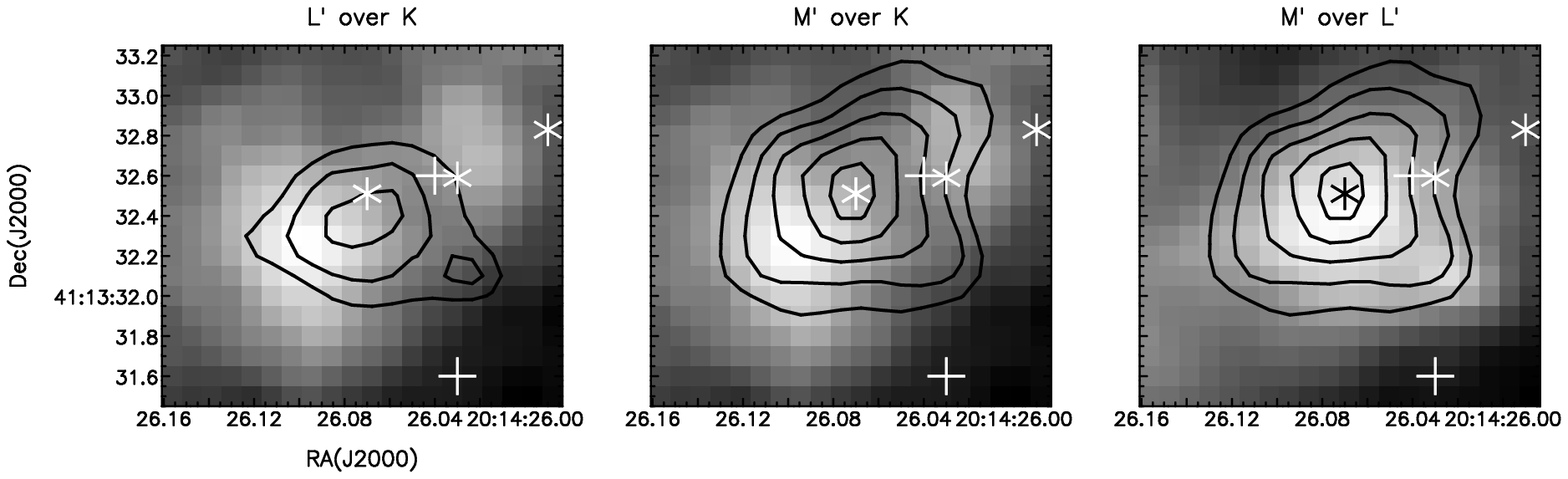}
\figcaption{\small K, L$^{\prime}$ and M$^{\prime}$ emission from the central region.
{\it left}: K-band image L$^{\prime}$-band emission contours. The asterisks mark 
water masers (Tofani et al 1995). The central cross mark locates the mm-wave emission 
(an average of the peak positions at 1.3, 3 and 7-mm), also coincident with one peak 
of the VLA 3.6-cm emission. The cross mark to the south is the location of the second 
3.6cm emission jet (Hofner et al 1999). {\it center}: K-band image with 
L$^{\prime}$-band contours.  {\it right}: L$^{\prime}$-band image with 
M$^{\prime}$-band contours.  }

\newpage
\begin{deluxetable}{rrrrrrrrrr}
\tablecolumns{10}
\tablewidth{0pt}
\tablecaption{Table 1 -- NIR Magnitudes, Colours and  Extinctions}
\tabletypesize{\small}
\tablehead{
\colhead{Object}      &    \multicolumn{3}{c}{Magnitude} &  \multicolumn{2}{c}{Colour} & \multicolumn{2}{c}{$A_{v,direct}$} &
\multicolumn{2}{c}{$A_{v,scattered}$}\\
\colhead{}      &\colhead{K} & \colhead{L$'$} & \colhead{M$'$} &       \colhead{K-L$'$}  &
\colhead{L$'$-M$'$}&
\colhead{(K-L$'$)}  & \colhead{(L$'$-M$'$)}  &  \colhead{(K-L$'$)}  & \colhead{(L$'$-M$'$)}   
}
\startdata
Whole     & 7.7     & 4.7  & 3.5      & 3.0     & 1.2      & 67       & 63       & 120    &  116  \\
SE lobe   & 13.5    & 10.8 & $\ldots$ & 2.7     & $\ldots$ & 60       & $\ldots$ & 113    & $\ldots$ \\
NW lobe   & 14.9    & 11.6 & $\ldots$ & 3.3     & $\ldots$ & 73       & $\ldots$ & 127    & $\ldots$ \\
Central   & $>$15.5 & 11.2 &  8.0     & $>$4.3  &  3.2     &   $>$96  & 168      & $>$149 & 221\\
Companion & $>$15.5 & 11.4 &  8.6     & $>$4.1  &  2.8     &   $>$91  & 148      & $>$144 & 200\\
\enddata
\tablecomments{The {\it direct} coulmns assumed extinguished star
light from stars with intrinsic K-L$'$ and L$'$-M$'$ colours of 0. The
{\it scattered} columns assumed  Rayleigh scattering before extinction.}
\end {deluxetable}

\end{document}